\documentclass[pra,aps,twocolumn,showpacs,10pt]{revtex4-1}
\usepackage{amssymb,amsmath,graphicx}

\begin{document}

\title{Quasi-ideal dynamics of vortex solitons embedded in top-hat nonlinear Bessel beams}

\author{Miguel A. Porras}
\affiliation{Grupo de Sistemas Complejos, ETSIME, Universidad Polit\'ecnica de Madrid, Rios Rosas 21, 28003 Madrid, Spain}
\author{Francisco Ramos}
\affiliation{Nanophotonics Technology Center, Universitat Polit$\grave{\textit{e}}$cnica de Val$\grave{\textit{e}}$ncia, Camino de Vera s/n, 46022 Valencia, Spain}

\begin{abstract}
Nonlinear Bessel beams in self-defocusing media are found to be the natural, non-diffracting background where vortex solitons can be nested, interact and survive for propagation distances that are one order of magnitude larger than in the usual Gaussian or super-Gaussian backgrounds. The dynamics of the vortex solitons approaches that in the ideal, uniform background, preventing vortex spiraling and decay, which eases vortex steering for applications.
\end{abstract}

\maketitle

\noindent An optical vortex soliton (OVS) is an intensity dip carrying a phase dislocation in a bright background field \cite{NEU90,DES05,SWAR92}. Even if the balance between diffraction and self-defocusing nonlinearity is stable in a single OVS, its propagation as such is always limited by the finiteness of the background beam where the OVS has to be inserted in any real setting. Self-defocusing accelerates diffraction spreading of the finite background \cite{SWAR92,TIKH96}, which severely limits the applications of OVSs and OVS arrays as, e. g., waveguides for other beams, particle and atom trapping or manipulation \cite{SWAR92,CARL00,KIV98,AND08}, soliton stabilization \cite{ADHI15}, and enhanced second harmonic generation \cite{SAL04}. Also, inhomogeneity of the intensity and phase of the background, typically a broad Gaussian or super-Gaussian (SG) beam, complicates the OVS dynamics and interactions of OVSs in arrays, which led to an intense study of OVS dynamics with the aim of steering their motion \cite{KIV98b,VELC98,ROZ00,ROZ97,NESH98}. These problems probably led to consideration of other configurations, such as ring-shaped (localized) OVSs \cite{DES05} in media with self-focusing or more complex self-defocusing nonlinearities \cite{BORIS1,BORIS2}, which can also guide other waves \cite{CID16}, but are limited to the single vortex of the ring soliton, excluding applications such as reconfigurable arrays of OVS-induced waveguides.

Surprisingly, embedding one or many OVSs in a nondiffracting beam in the self-defocusing medium has not been considered before, to our knowledge. These beams indeed exist in the form of fundamental nonlinear Bessel beams (NBBs) \cite{JOHA03}, and high-order NBBs, not described previously in self-defocusing media, since most of research focused on positive Kerr nonlinearity \cite{POR1,POR2,POR3,CAR16,ARN08}. As shown here, these beams may present arbitrarily wide regions of uniform intensity and phase that propagate without any change for long distances, only limited by instability effects.
We show that NBBs with wide "platform" can be generated placing a standard Bessel beam generator in front of the nonlinear medium.
We then consider OVSs nested in the platform of fundamental NBBs to show that they survive as OVSs and interact closely approaching the much simpler dynamics in the uniform background, for distances that are one order of magnitude larger than in standard backgrounds of similar size. E. g., a single off-axis OVS can remain undistorted and at rest, and an OVS pair rotate uniformly more than $360^\circ$ in a circular orbit, in about $70$ nonlineal lengths, approximately ten times the distances and angles reported in \cite{NESH98}, and four times the angles in \cite{ROZ00}. Our analysis also offers a unified view of the nature of the OVS-background system as a particular vortex-carrying conical beam in the self-defocusing medium, offering new possibilities for their generation.

As is well-known, OVSs are solutions of the form $A=\sqrt{I_0}\,b(r)e^{im\varphi} e^{i\delta z}$ to the nonlinear Schr\"odinger equation (NLSE)
\begin{equation}\label{NLSE}
\partial_z A = \frac{i}{2k}\Delta_\perp A + i \frac{kn_2}{n}|A|^2 A\, ,
\end{equation}
in a medium with $n_2<0$ \cite{DES05}. In the above equations, $(r,\varphi,z)$ are cylindrical coordinates, $\Delta_\perp =\partial^2_r +(1/r)\partial_r +(1/r^2)\partial^2_\varphi$ is the transversal Laplacian operator, $A$ is the complex envelope of the monochromatic light beam $E=A\exp[-i(\omega t-kz)]$ of angular frequency $\omega$ and propagation constant $k = (\omega/c)n$ in a medium of linear refractive index $n$. The axial wavenumber is shortened by $\delta<0$, related to its background intensity by $I_0=n|\delta|/k|n_2|$, and the radial profile verifies $b(r)\rightarrow 1$ as $r\rightarrow \infty$ and $b(r)\sim r^{|m|}$ as $r\rightarrow 0$, where the vortex of topological charge $m=\pm 1,\pm 2,\dots$ is located.

\begin{figure}[t!]
\centering
\includegraphics*[width=2.9cm]{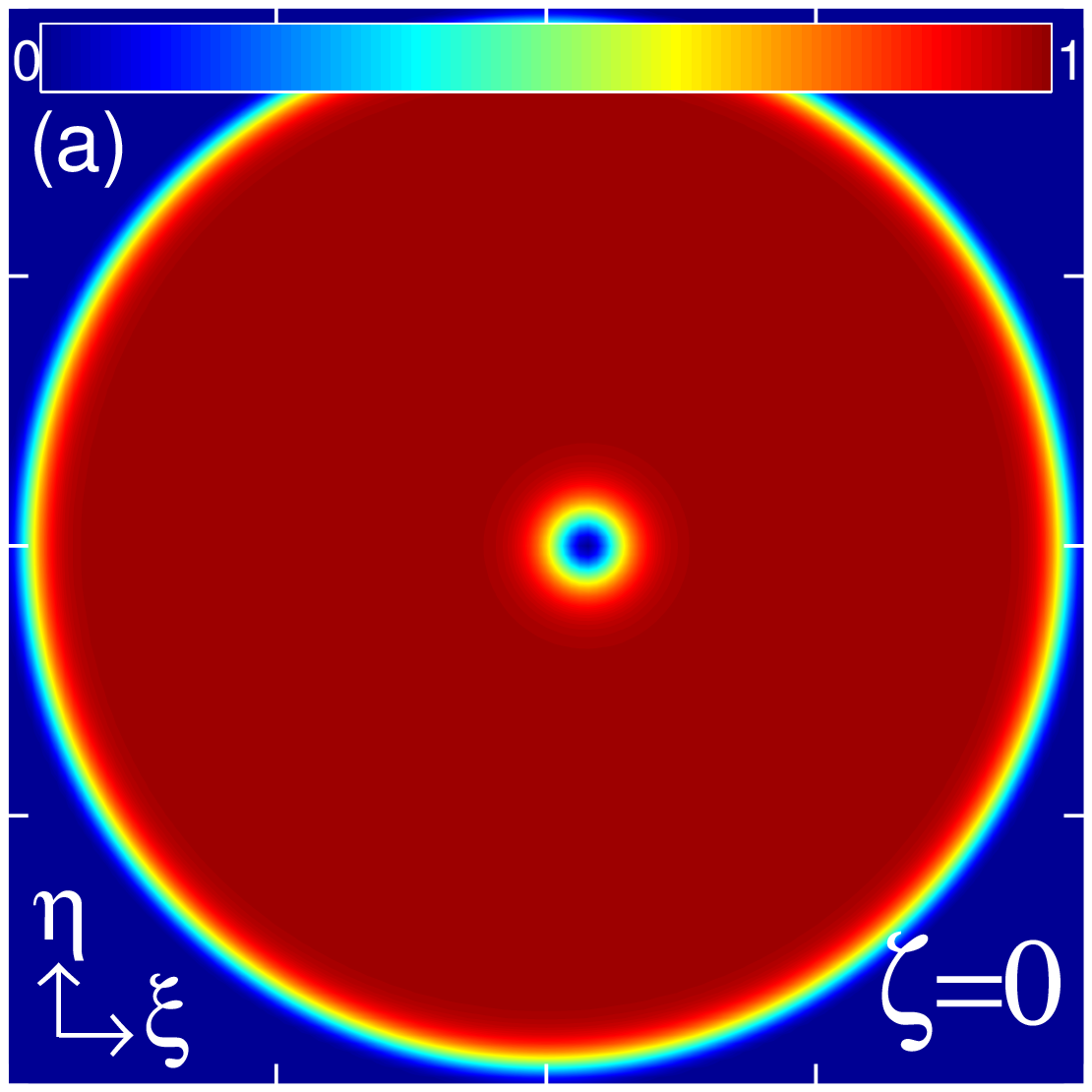}\includegraphics*[width=2.9cm]{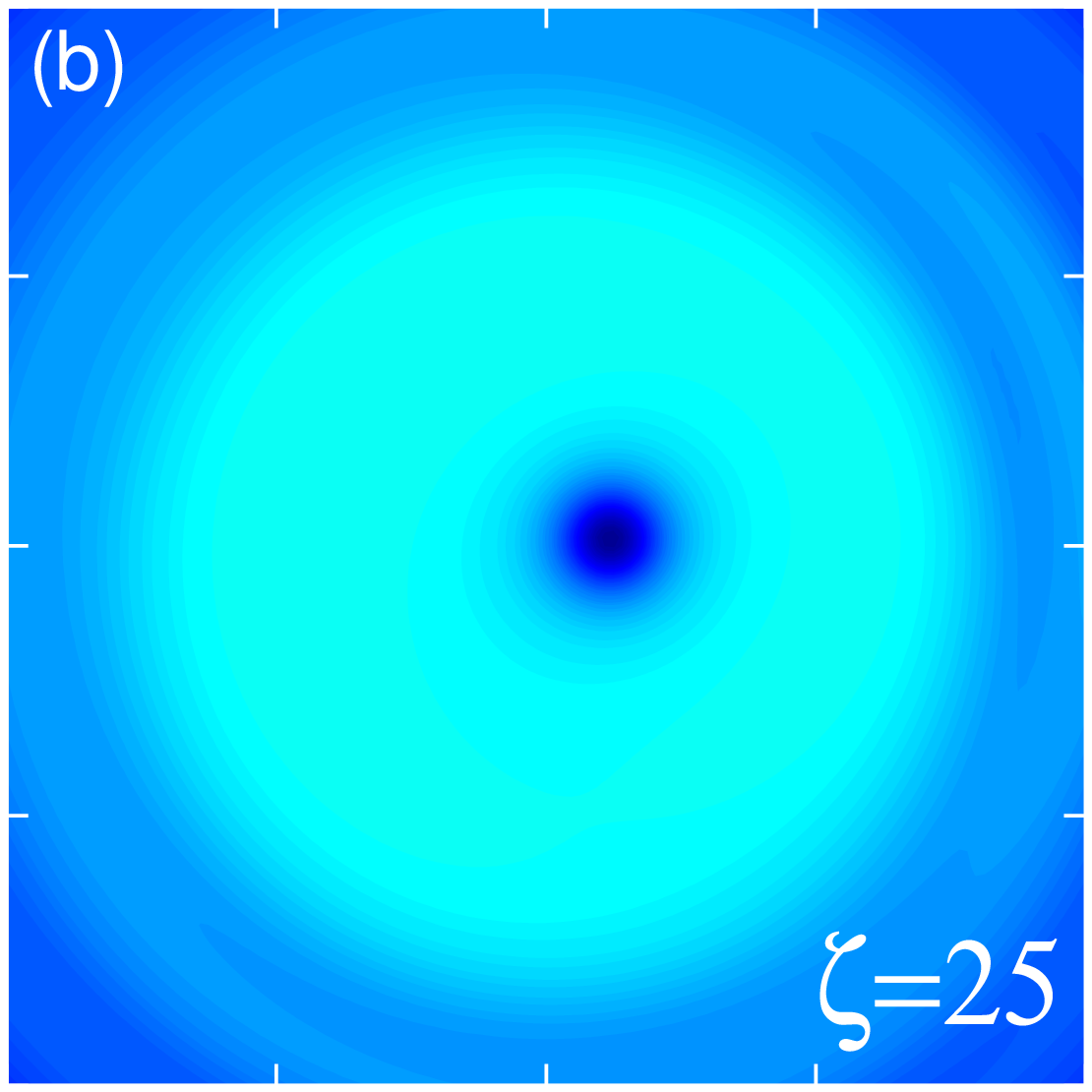}\includegraphics*[width=2.9cm]{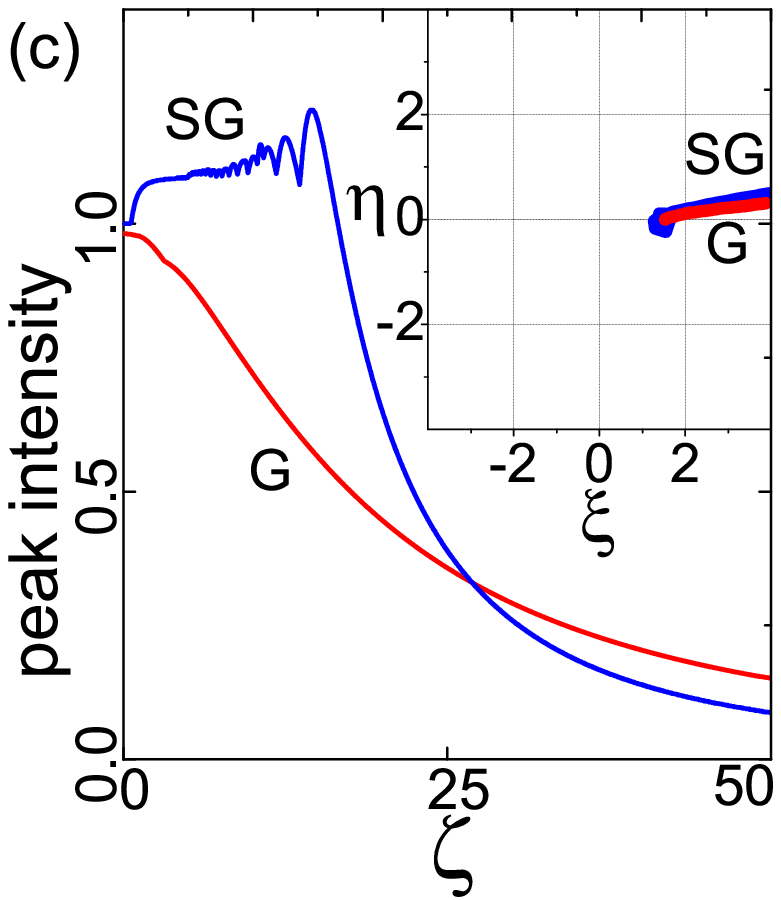}
\caption{\label{Fig1} (a,b) A singly-charged OVS ($m=1$) initially displaced $1.5$ from the center of the SG background $u_{\rm BG}=\exp[-(\rho/19.7)^{40}]$ at the indicated distances. Normalized Cartesian coordinates are $(\xi,\eta)=\sqrt{k|\delta|}(x,y)$.
The distance between axis tics is $10$. (c) Peak intensity as a function of propagation distance for the above SG beam with the nested OVS and for the Gaussian beam $u_{BG}=\exp[-(\rho/32.6)^2]$, of the same ${\rm FWHM}=38.4$ and with the same nested OVS. Inset in (c): trajectory of the OVS in both cases.}
\end{figure}

We find it convenient to study OVSs and their dynamics using the dimensionless coordinates and envelope $\zeta=|\delta|z$, $\rho=\sqrt{k|\delta|} r$ and $u=\sqrt{k|n_2|/n|\delta|}A$. For comparison purposes, the characteristic nonlinear length $L_{\rm NL}=1/k|n_2|I_0$ corresponds to $\zeta_{\rm NL}=1$. The NLSE (\ref{NLSE}) in the dimensionless form reads
\begin{equation}\label{NLSE2}
\partial_\zeta u = \frac{i}{2}\Delta_\perp u- i|u|^2 u \,,
\end{equation}
where now $\Delta_\perp =\partial^2_\rho +(1/\rho)\partial_\rho +(1/\rho^2)\partial^2_\varphi$, and the OVS as $u_m=b_m(\rho)e^{im\varphi}e^{-i\zeta}$, where the exact radial profile $b_m(\rho)$ of the OVS is determined by
\begin{equation}\label{EST}
\frac{d^2 b}{d\rho^2}+ \frac{1}{\rho}\frac{db}{d\rho}- \frac{m^2}{\rho^2} b +2b -2b^3 =0\, ,
\end{equation}
subjected to the boundary conditions $b(\rho)=C_{|m|}\rho^{|m|}$ as $\rho\rightarrow 0$ and $b(\rho)\rightarrow 1$ as $\rho\rightarrow\infty$. These conditions are satisfied for the specific values $C_1\simeq 0.824754$, $C_2\simeq 0.306198$,\dots. The corresponding radial profiles can be approached by $b_m\simeq [\tanh(C_{|m|}^{1/|m|}\rho)]^{|m|}$. In Ref.\cite{ROZ97}, $b_1(\rho)\simeq\tanh(0.787\rho)$ is used because it gives a better overall fitting (not only at $\rho\rightarrow 0$).

\begin{figure}[t!]
\centering
\includegraphics*[width=2.9cm]{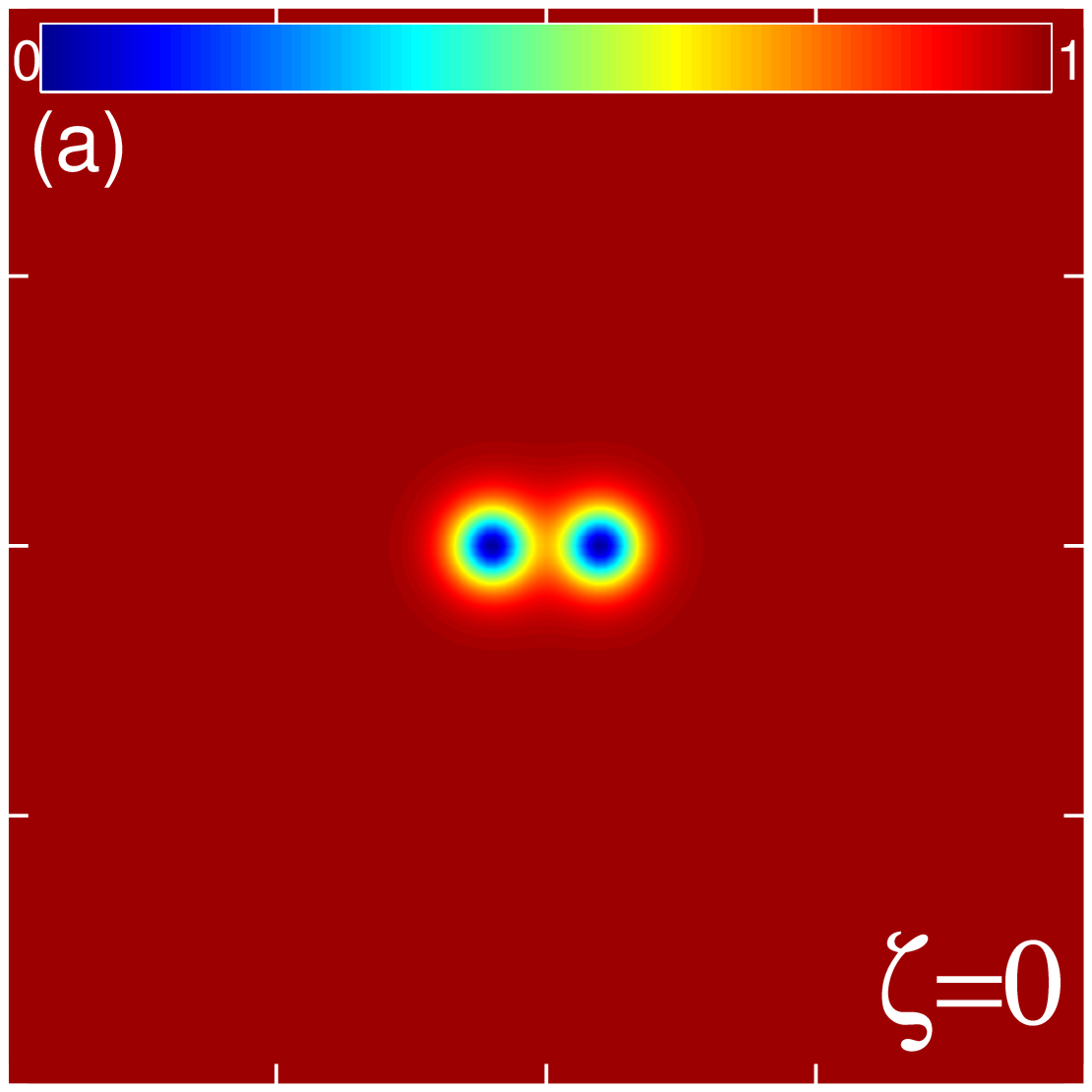}\includegraphics*[width=2.9cm]{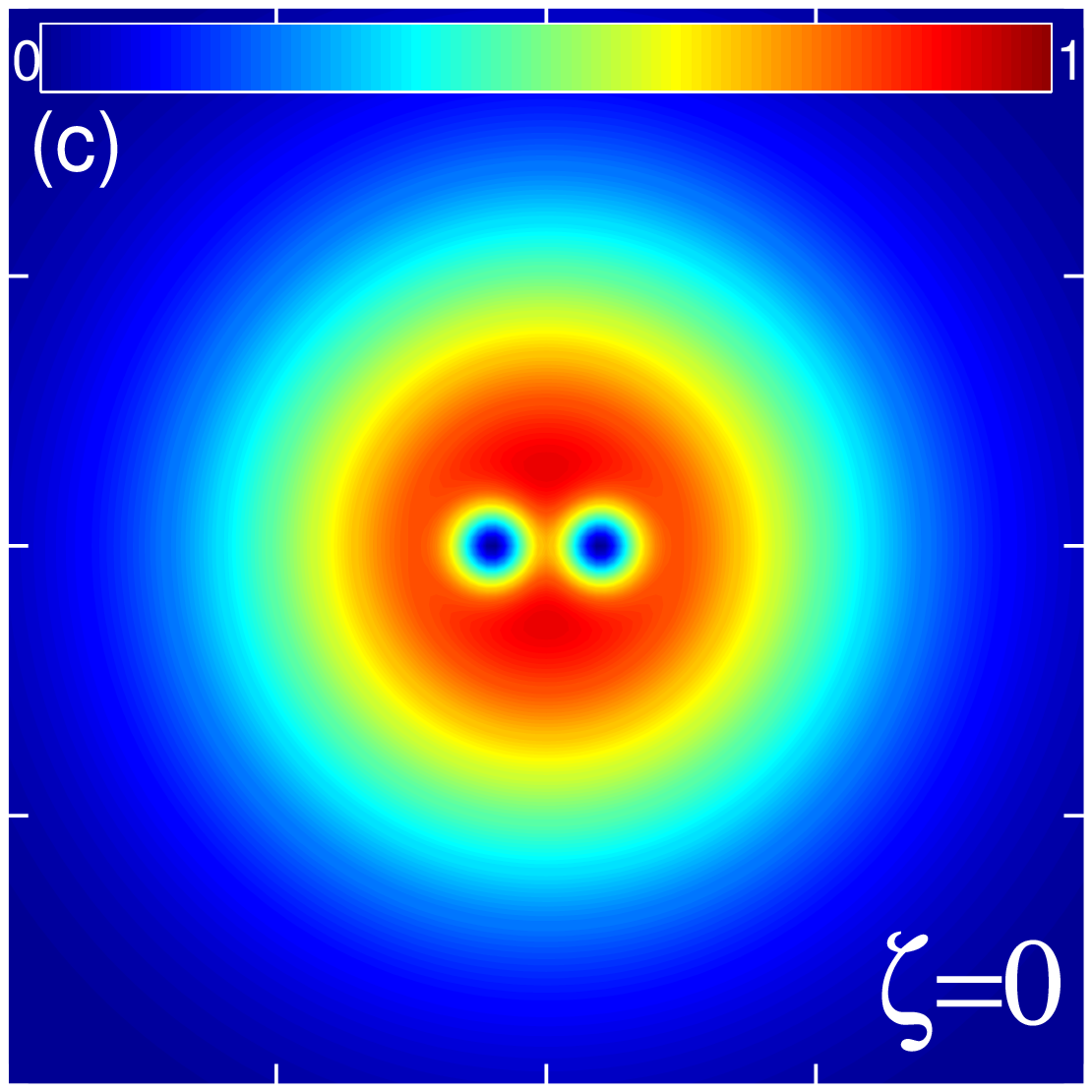}\includegraphics*[width=2.9cm]{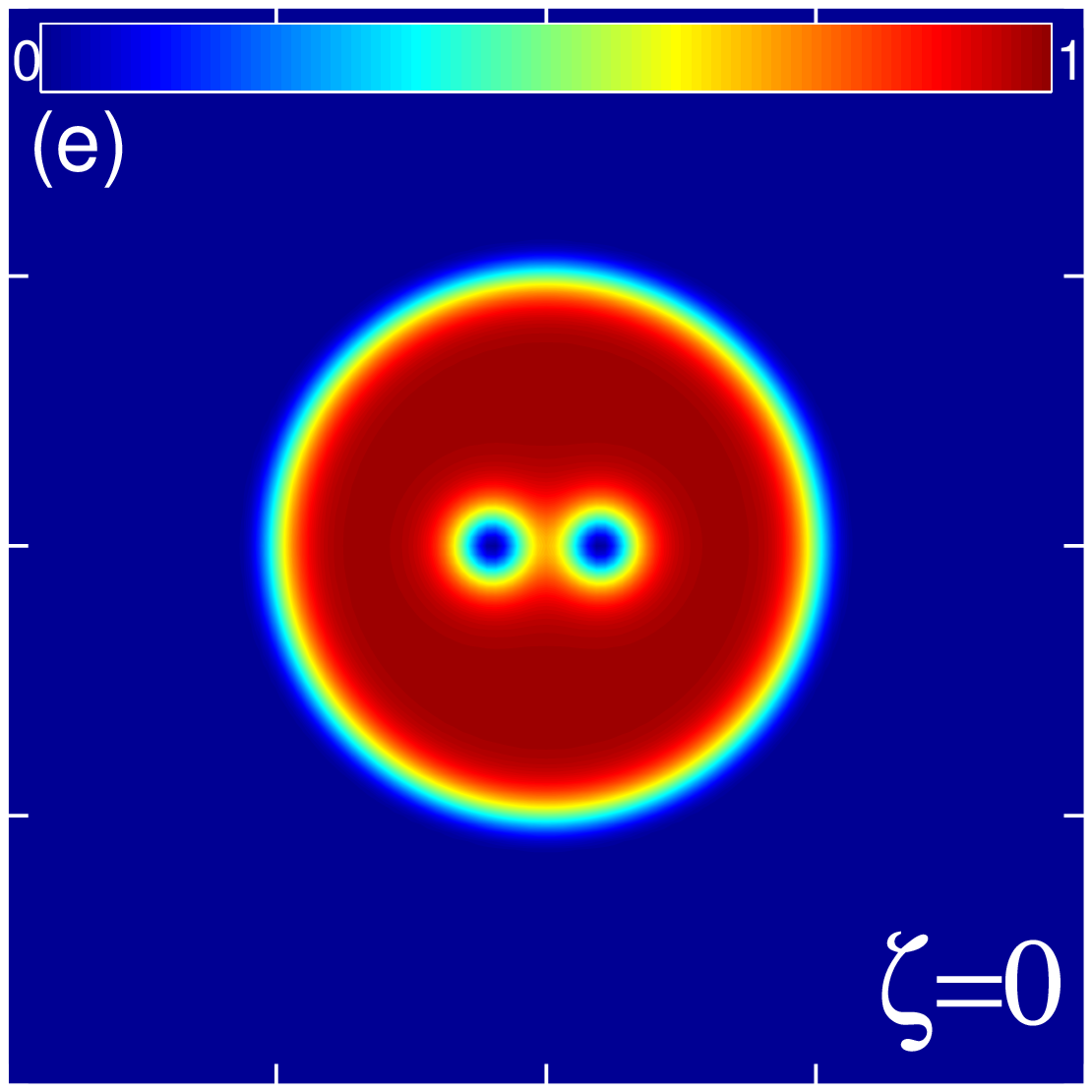}
\includegraphics*[width=2.9cm]{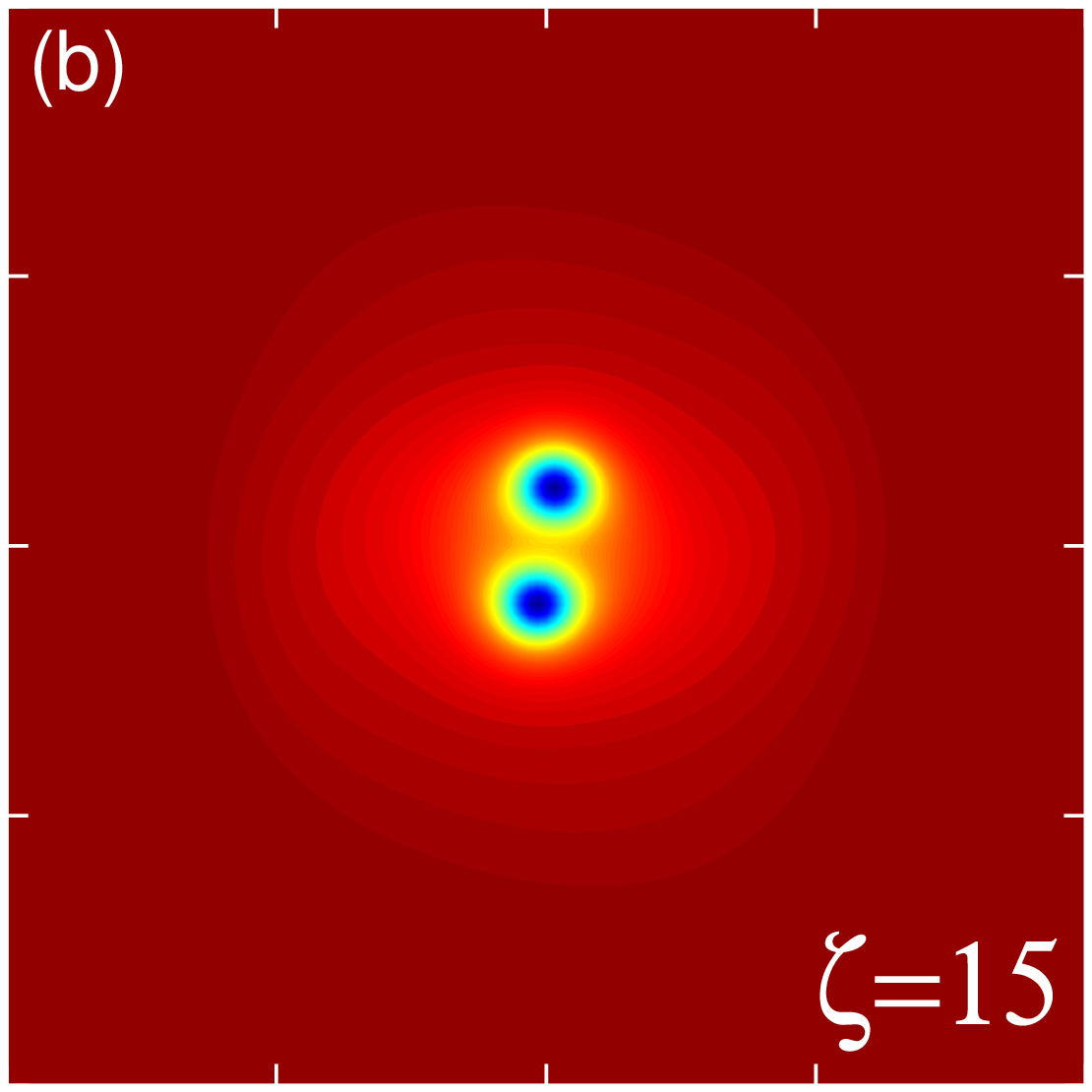}\includegraphics*[width=2.9cm]{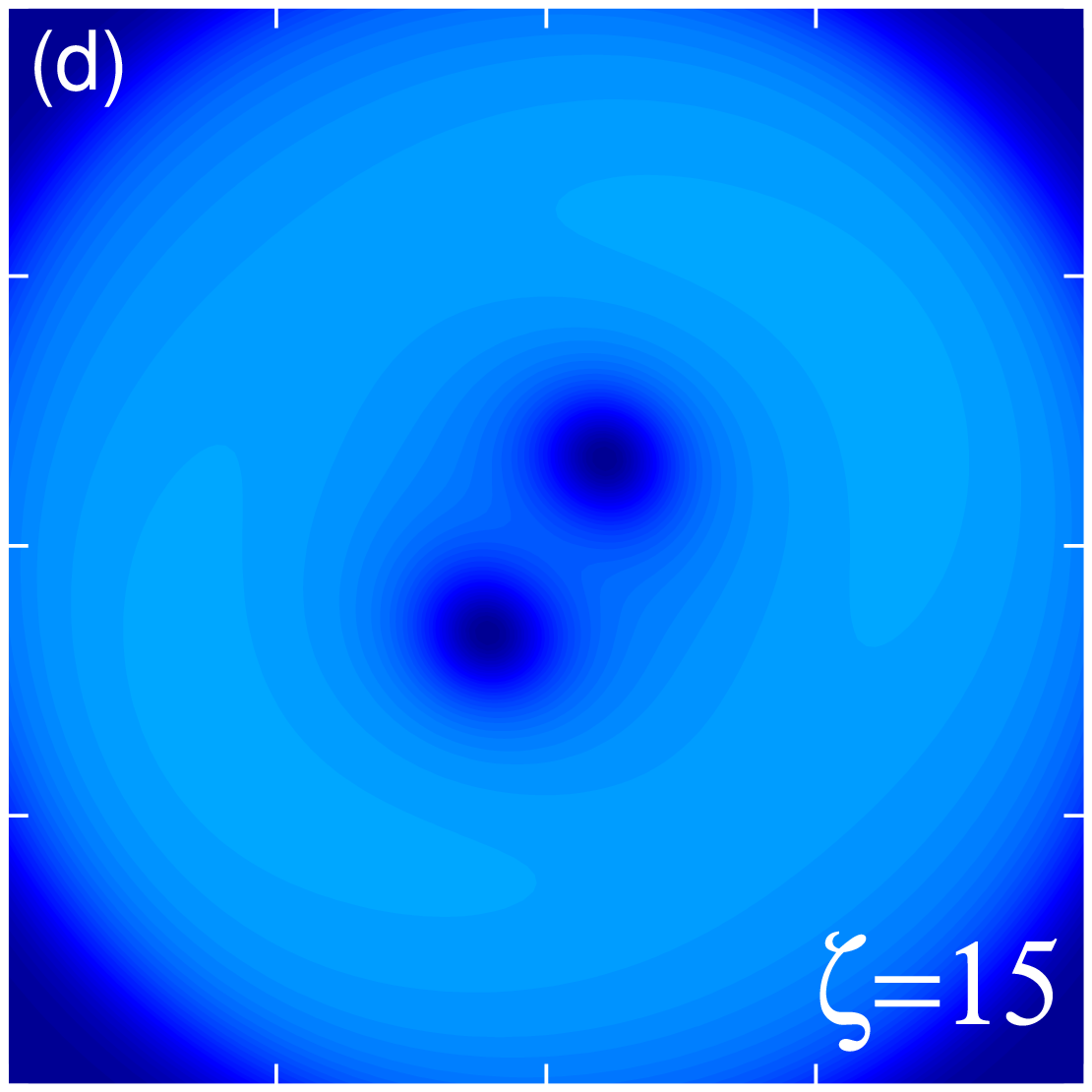}\includegraphics*[width=2.9cm]{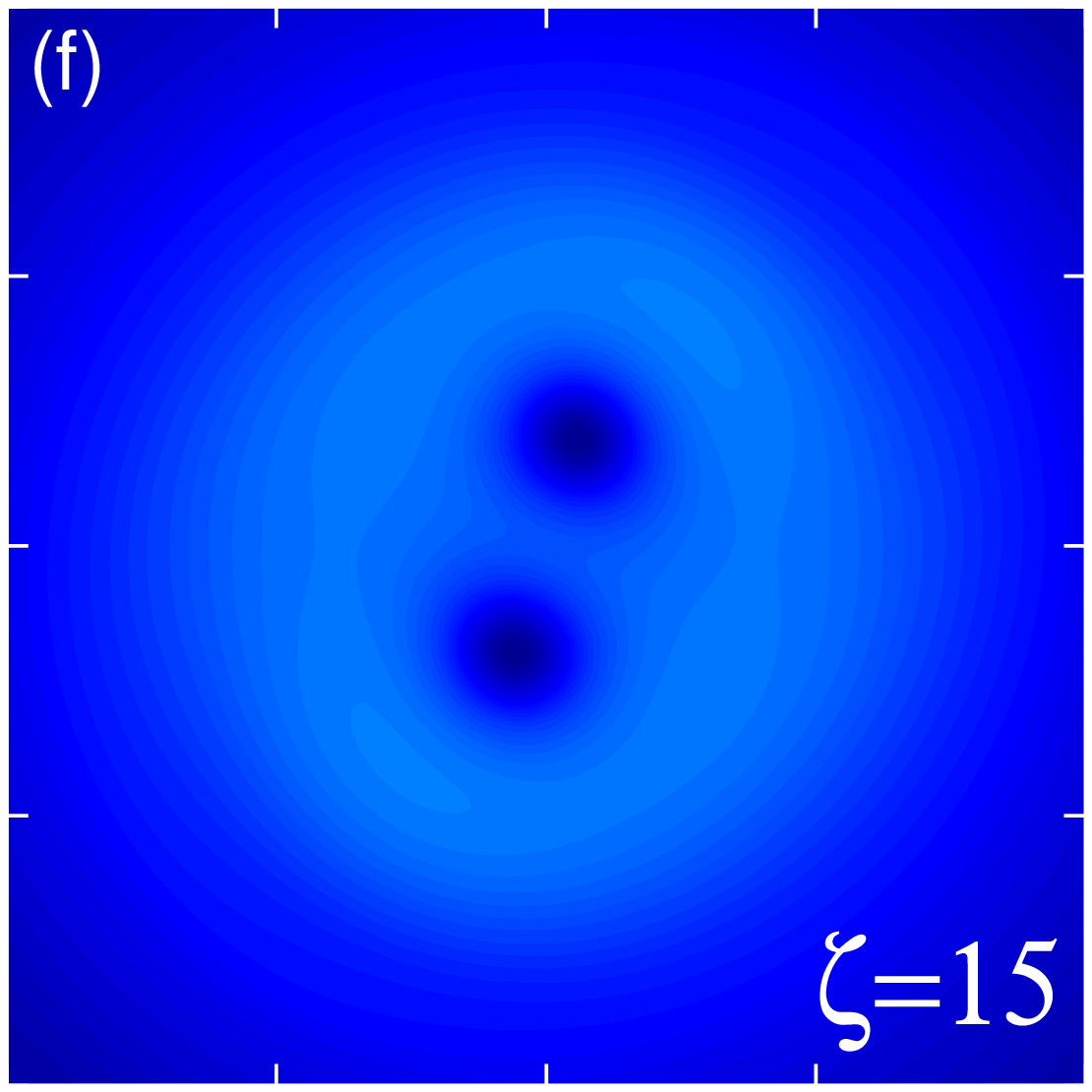}
\includegraphics*[height=2.8cm]{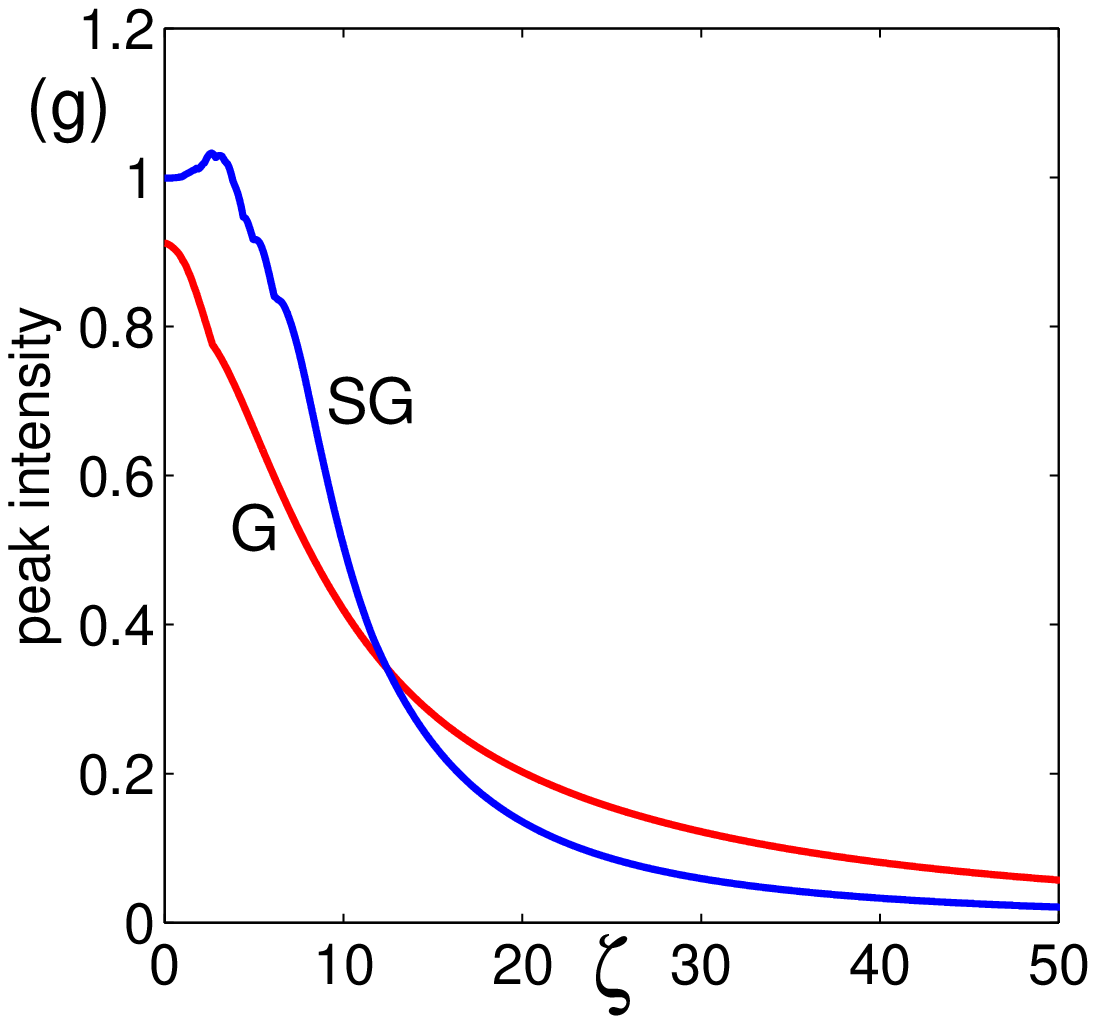}\includegraphics*[height=2.8cm]{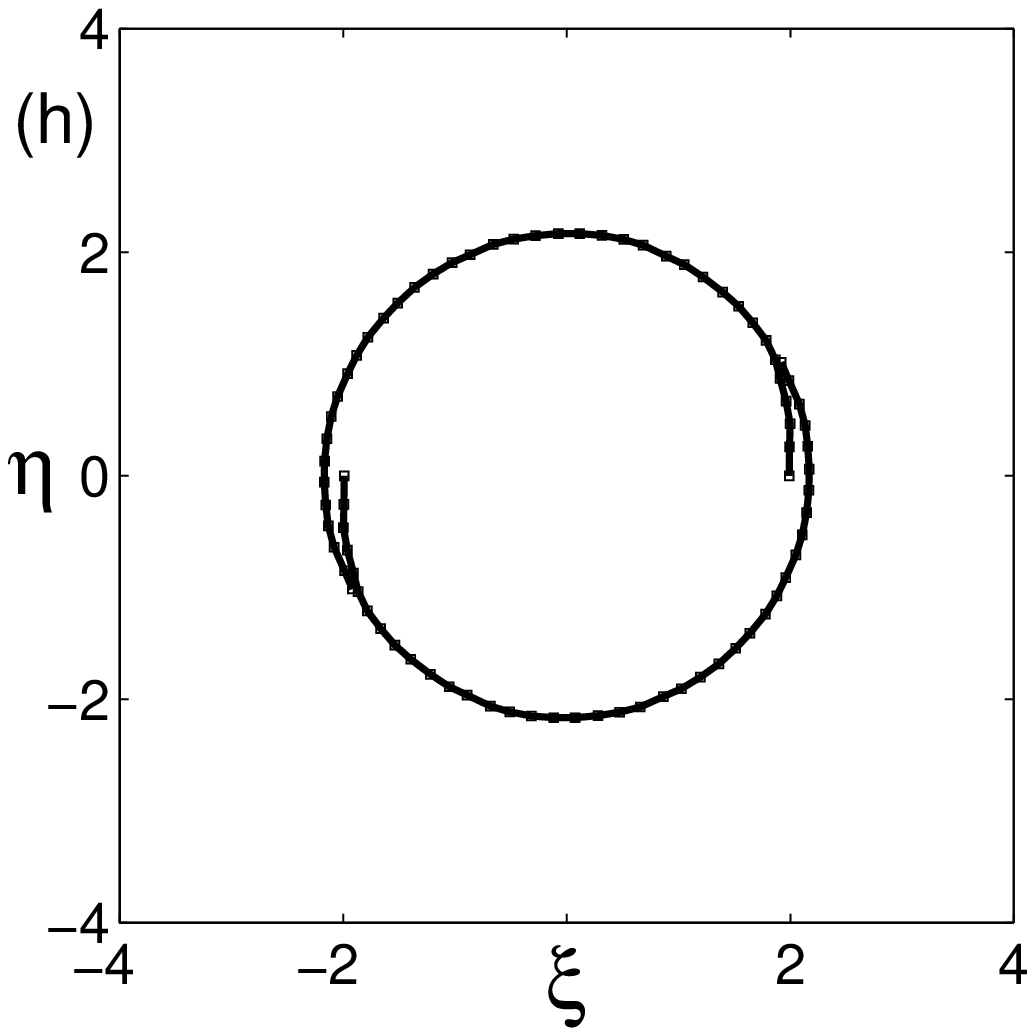}\includegraphics*[height=2.8cm]{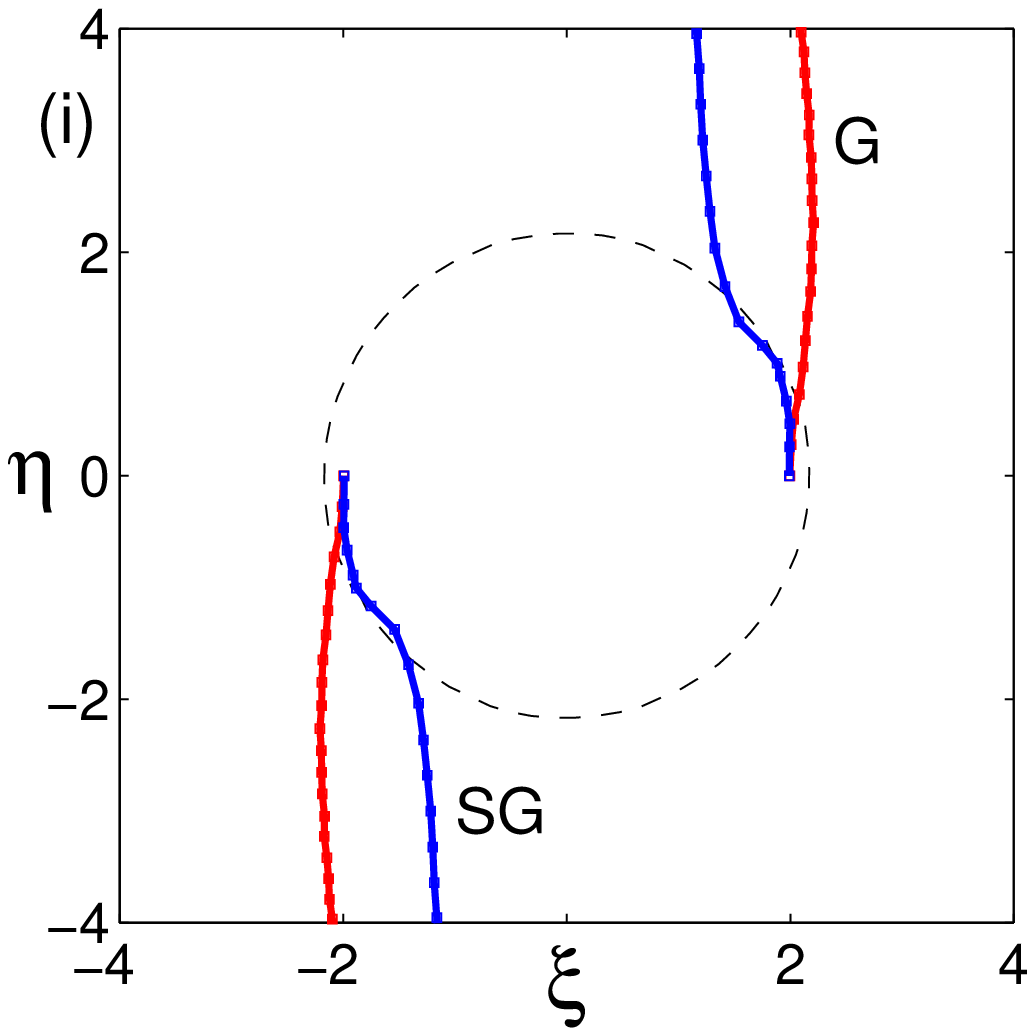}
\caption{\label{Fig2} Two singly-charged OVSs separated $d=4$ in (a,b) uniform (c,d) Gaussian $u_{BG}=\exp[-(\rho/16.6)^2]$ and (e,f) SG $u_{\rm BG}=\exp[-(\rho/10.7)^{15}]$ backgrounds, at the indicated distances. Normalized horizontal and vertical axes and distance between tics are as in Fig. \ref{Fig1}. (g) Peak intensity as a function of propagation distance for the Gaussian and SG beams with the nested vortices. (h,i) Ideal trajectories in the uniform background and in the Gaussian and SG backgrounds.}
\end{figure}

Ideally, OVSs need an unlimited background to subsist. In practice, OVSs are usually nested in a broad Gaussian beam, or in a SG beam in order to simulate the flat background \cite{TIKH96,VELC98,NESH98}, where they undergo a complex individual or interaction dynamics governed by their topological charges, mutual disposition, and the phase and intensity gradients of the background \cite{KIV98b,VELC98,ROZ00,ROZ97,NESH98}. In either case, OVSs are eventually dispersed and destroyed when joint action of diffraction and self-defocusing expand the background. In our examples, the initial condition in the NLSE (\ref{NLSE2}) is of the type $u(\rho,\varphi,\zeta=0)= u_{\rm BG} \prod_{j=1}^N u_{m_j}$, where the OVSs $u_{m_j}$ are placed at different positions in the background field $u_{\rm BG}$. For example, the single vortex slightly displaced from the Gaussian or SG center in Fig. \ref{Fig1} widens as the background intensity diminishes [Figs. \ref{Fig1}(a,b)] and moves almost radially outwards [in set of Fig. \ref{Fig1}(c)]. In contrast, the single OVS would subsist forever at rest in the infinite background. As a second example, Fig. \ref{Fig2} illustrates the interaction dynamics of two equal OVSs nearby in uniform, Gaussian and SG backgrounds. The two OVSs broaden and spiral-out as the background intensity diminishes, while the vortices in the ideal background simply rotate indefinitely at constant angular velocity following circular trajectories [aside the initial transient of reshaping from  $b_1(\rho)\simeq\tanh(0.787)$) to the exact 2-OVS structure seen in Fig. \ref{Fig2}(h)]. In these examples, the FWHMs of the backgrounds compared to the OVSs size are similar to those in Refs. \cite{ROZ97,NESH98}.

We consider the possibility of nesting the OVSs in the propagation-invariant NBBs supported by the self-defocusing medium. Unlike the Kerr-compressed, highly unstable NBBs in transparent media with $n_2>0$ \cite{JOHA03,POR4}, NBBs with $n_2<0$ may have arbitrarily wide and flat, intensity profiles and their instability is weaker. In addition, there is an intimate connection between these NBBs and OVSs. NBBs are solutions to (\ref{NLSE}) of the same form as OVSs, i. e., $A=\sqrt{I_0}\,b(r)e^{im\varphi} e^{i\delta z}$, with $\delta<0$ too, but $b(r)$ approaches zero at large radius as $b_{\infty}J_m(k\theta r)$, where $\theta=\sqrt{2|\delta|/k}$ is the cone angle, and $b_{\infty}$ is a constant. Therefore $I_0=n|\delta|/k|n_2|$ has not the meaning of a background intensity for NBBs. In our dimensionless variables, NBBs read $u=b(\rho)e^{im\varphi}e^{-i\zeta}$, with $b(\rho)$ satisfying also (\ref{EST}) with the boundary condition $b(\rho)=C\rho^{|m|}$ as $\rho\rightarrow 0$, but approaching zero as $b_\infty J_m(\sqrt{2}\rho)$. These solutions indeed exist for $0<C<C_0\equiv 1$ for $m=0$, and for $0<C<C_{|m|}$ for each given charge of the vortex at the NBB center $\rho=0$.

\begin{figure}
\begin{center}
\includegraphics*[width=4.2cm]{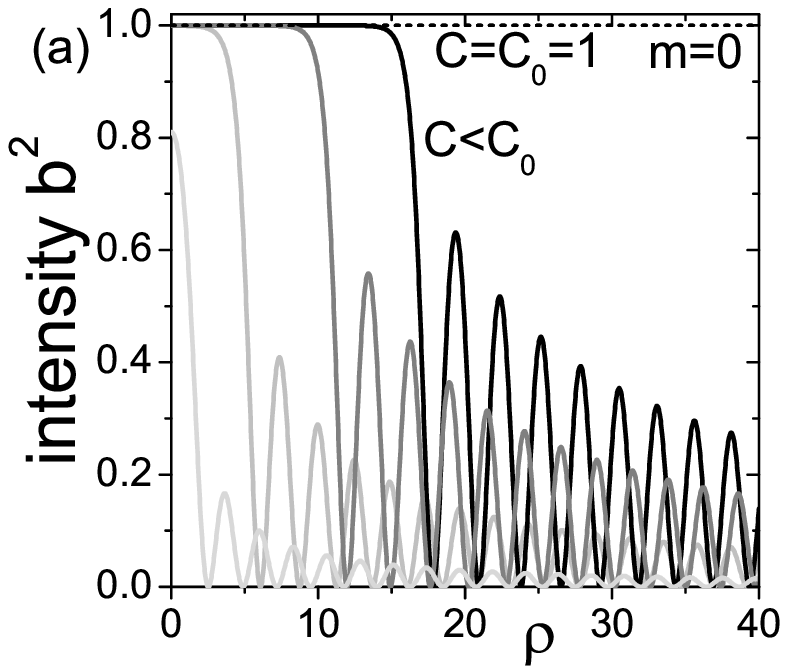}\includegraphics*[width=4.2cm]{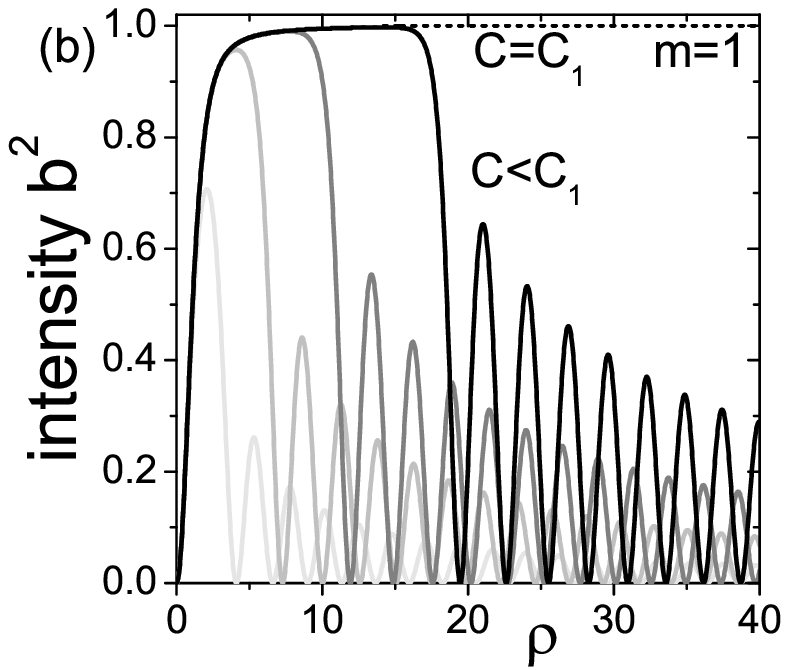}\\
\includegraphics*[width=4.2cm]{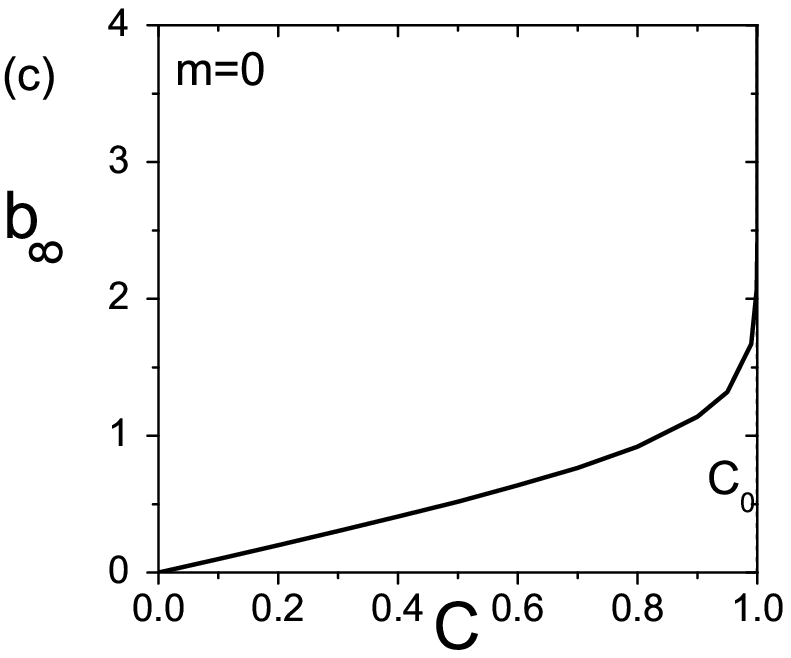}\includegraphics*[width=4.2cm]{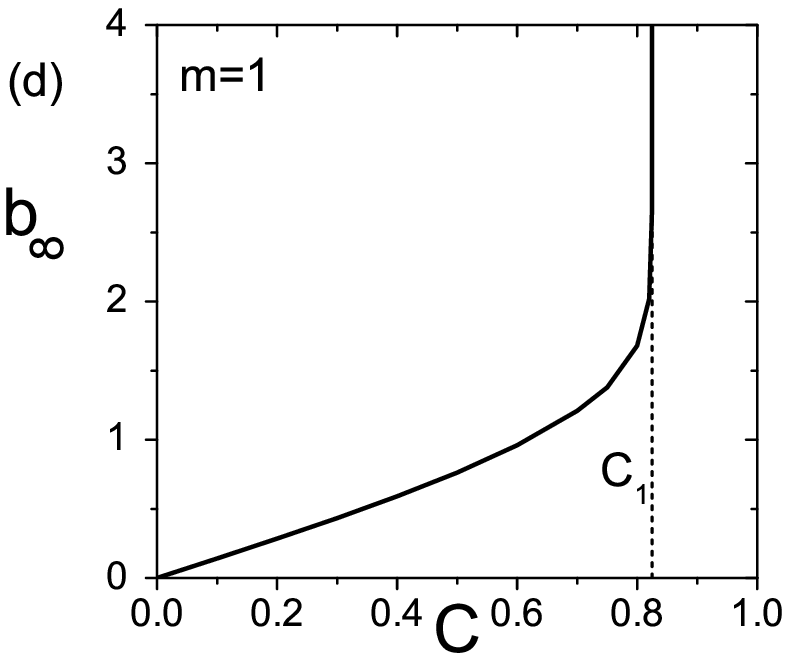}\\
\includegraphics*[width=4.2cm]{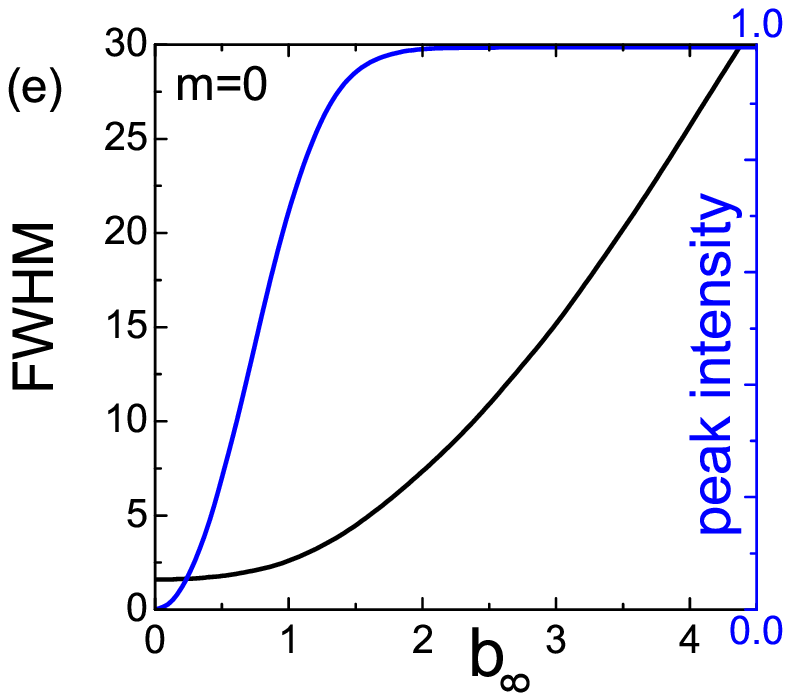}\includegraphics*[width=4.2cm]{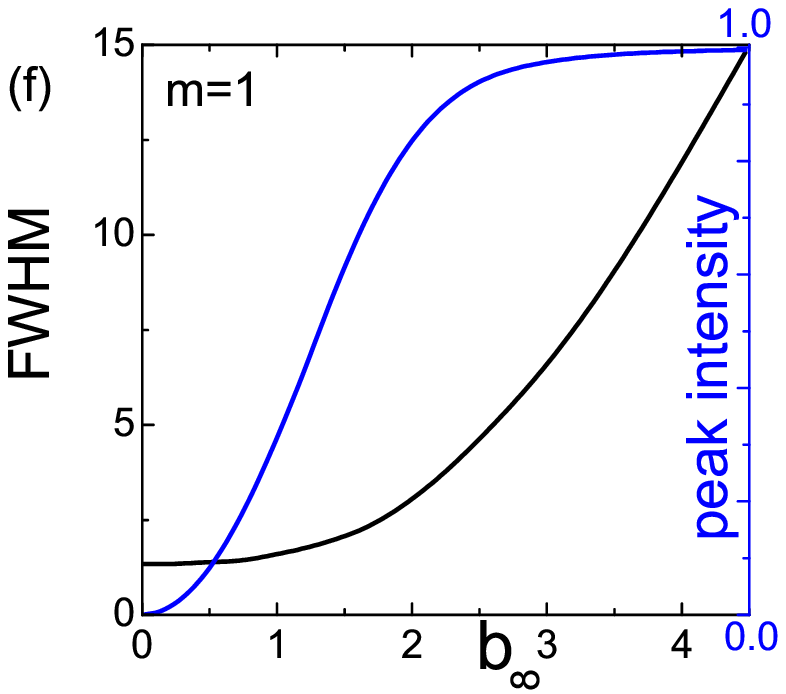}
\end{center}
\caption{\label{Fig3} (a,b) Radial intensity profiles of NBBs with increasing $C$ featuring increasingly wide region of flat unit intensity. The limit $C\rightarrow C_{|m|}$ is the uniform nonlinear plane wave (for $m=0$) and the OVS (for $|m|>0$), shown as dashed curves. (c,d) Amplitude $b_\infty$ of the linear Bessel tail as a function of $C$. (e,f) FWHM of the central maximum or inner ring (black curves) and its intensity (blue curves) as functions of $b_\infty$.}
\end{figure}

The limit of low $C$ is the linear Bessel beam. As $C$ increases, the intensity profile features wider and flatter central maximum of dimensionless intensity very close to unity (intensity $I_0$) for the fundamental NBB ($m=0$), and wider and flatter inner ring of intensity also close to unity for high-order NBBs ($|m| > 0$) [Figs. \ref{Fig3}(a) and (b)]. The uniform nonlinear plane wave $b(\rho)=1$ for $m=0$, and the OVS of the corresponding charge $m\neq 0$ correspond to the limits $C\rightarrow C_{|m|}$ [dashed curves in Figs. \ref{Fig3}(a) and (b)]. Thus, for NBBs, $I_0$ is the maximum attainable intensity, given cone angle. The OVS of topological charge $m\neq 0$ is the limiting NBB of the same charge as the inner ring becomes infinetely wide. Vice versa, the NBB with $m\neq 0$ and wide inner ring can be considered as the natural, finite background where the vortex in its center can subsist for long distances, only limited by the instability effects considered below.

Other relevant properties of NBBs are shown in Figs. \ref{Fig3}(c-f). The amplitude $b_\infty$ of the linear Bessel tails, $b_\infty J_m(\sqrt{2}\rho)$, as obtained from their numerical profiles, and are seen in Figs. \ref{Fig3}(c) and (d) to grow monotonously from zero up to infinity with $C$ increasing from $0$ to $C_{|m|}$. As seen below, $b_\infty$ is a directly accessible parameter in the proposed experimental generation of NBBs. It is then more practical to specify a NBB by the value of $b_\infty$, ranging from $0$ to $\infty$, than by $C$ in $[0,C_{|m|}]$. Figures \ref{Fig3}(e) and (f) show that the FWHM of the central maximum of intensity (for $m=0$), or of the inner ring (for $m=1$), and the peak intensity, grows without bound and stabilizes in unity, respectively, as $b_\infty$ increases ($C$ approaches $C_{|m|}$).

\begin{figure}
\begin{center}
\includegraphics*[width=4.5cm]{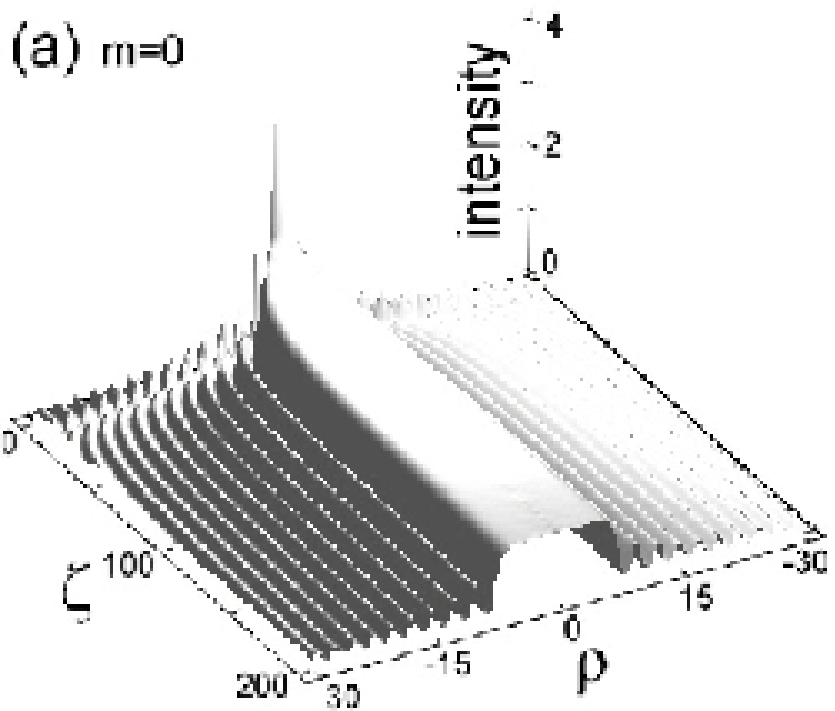}\includegraphics*[width=4.5cm]{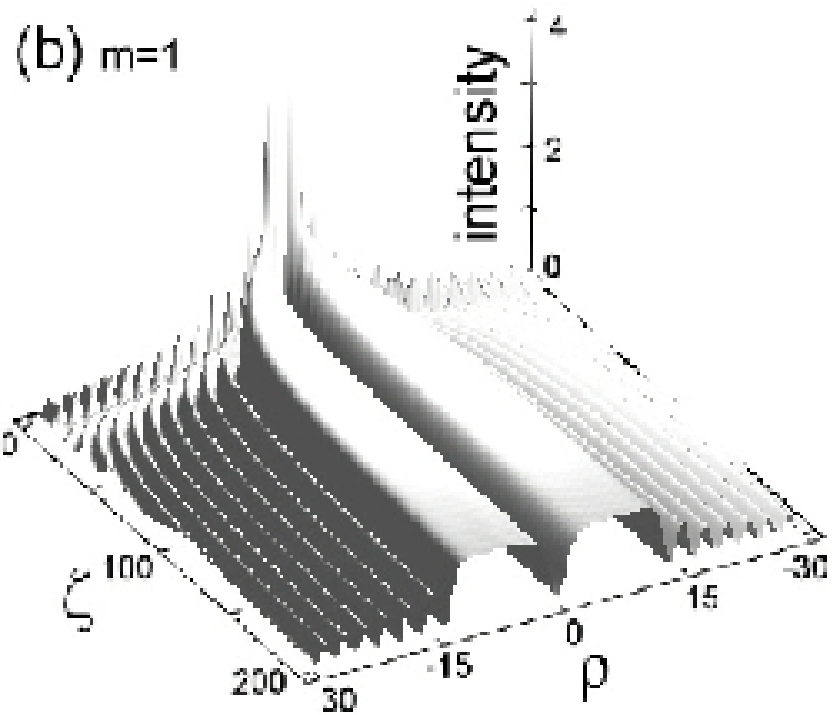}
\end{center}
\caption{\label{Fig4} Variation of the radial intensity profile with propagation distance in a self-defocusing medium for linear Bessel beams $b_\infty J_m(\sqrt{2}\rho)\exp(im\varphi)$ launched in the medium. In (a), $m=0$ and $b_\infty=3.2$, leading to the formation of the fundamental NBB of unit peak amplitude and FWHM of about $15$, as predicted in Fig. \ref{Fig3}(c). In (b), $m=1$ and $b_\infty=4.1$, leading to the formation of a vortex NBB of unit peak amplitude and FWHM of the first ring of about $12.5$, as predicted by Fig. \ref{Fig3}(c).}
\end{figure}

Figure \ref{Fig4} illustrates how NBBs with wide platform can be generated. Both in Fig. \ref{Fig4}(a) for $m=0$ and (b) for $m=1$ the initial condition in the NLSE (\ref{NLSE2}) is the linear Bessel beam $b_\infty J_m(\sqrt{2}\rho)\exp(im\varphi)$ of the amplitude $b_\infty$. Extrapolating the results in Ref. \cite{POR3} for self-focusing media to self-defocusing media, the NBB that is spontaneously formed from the input linear Bessel beam at long enough propagation distances is that preserving the amplitude of the linear Bessel tail, $b_\infty$. Therefore, the width of the platform of the NBB can be predicted from the amplitude $b_\infty$ of the input linear Bessel beam and Figs. \ref{Fig3}(e) and (f) for fundamental and high-order NBBs. According also to Figs. \ref{Fig3}(e) and (f), the higher the intensity of the input linear Bessel beam, the wider the platform of the formed NBB.

\begin{figure}[t!]
\centering
\includegraphics*[width=2.9cm]{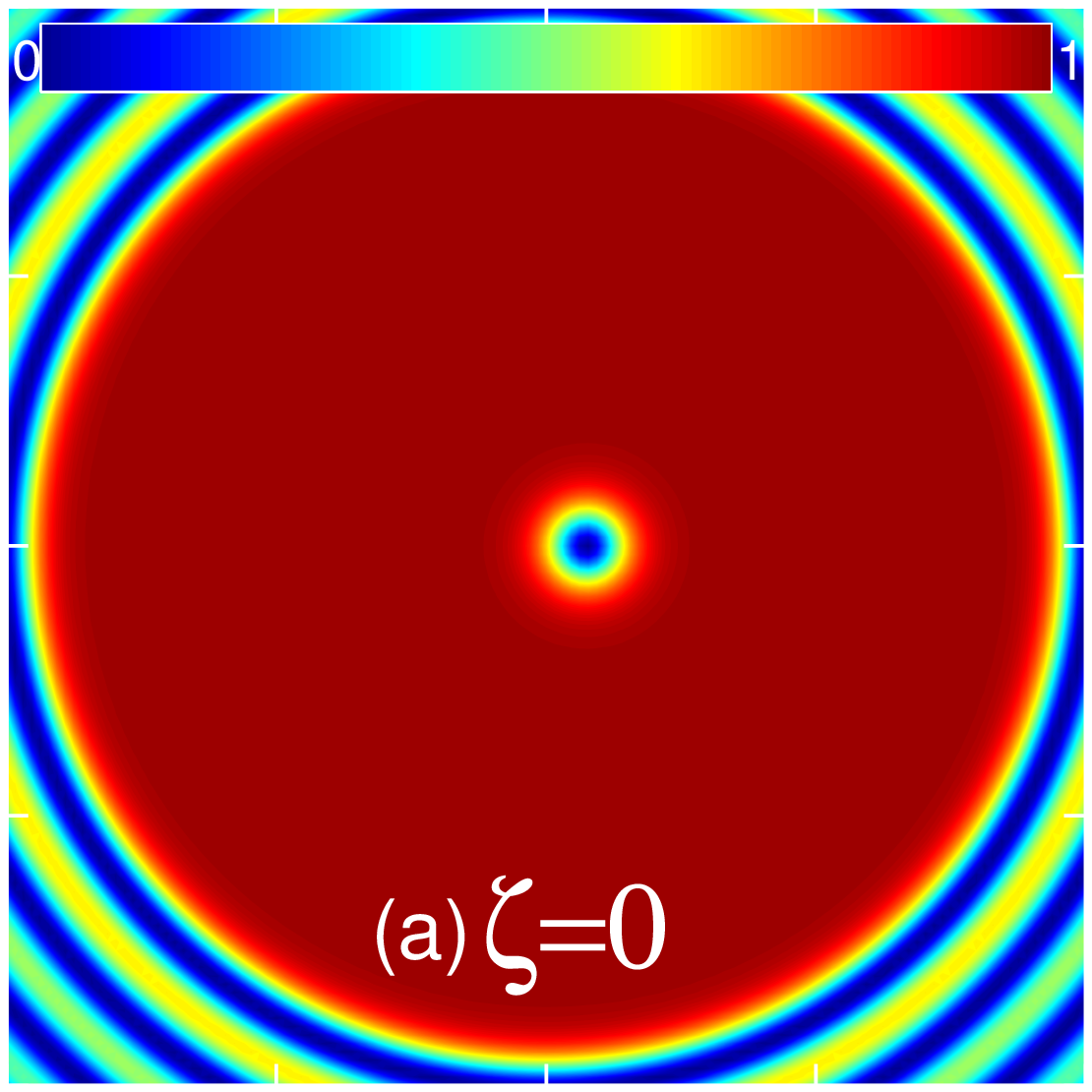}\includegraphics*[width=2.9cm]{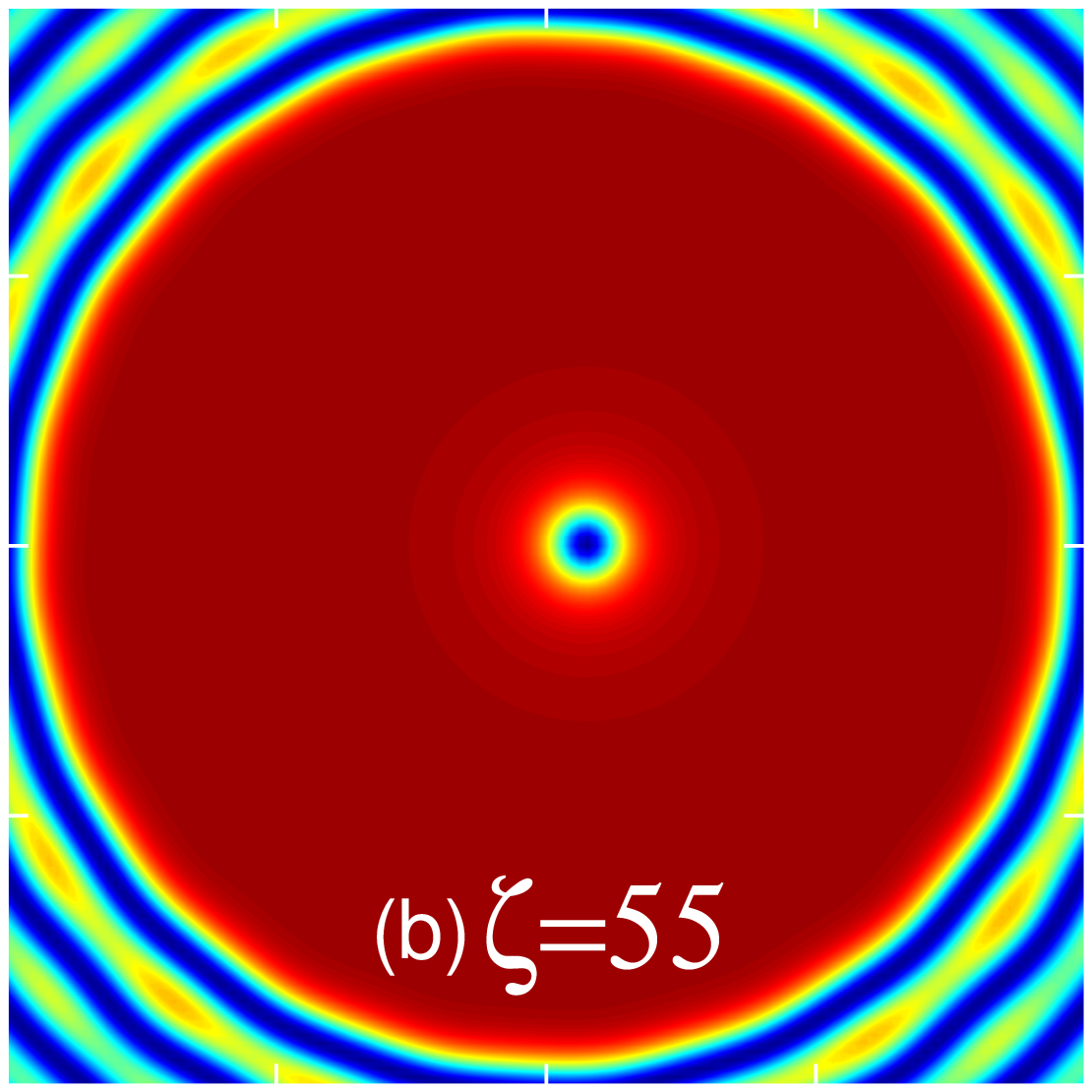}\includegraphics*[width=2.9cm]{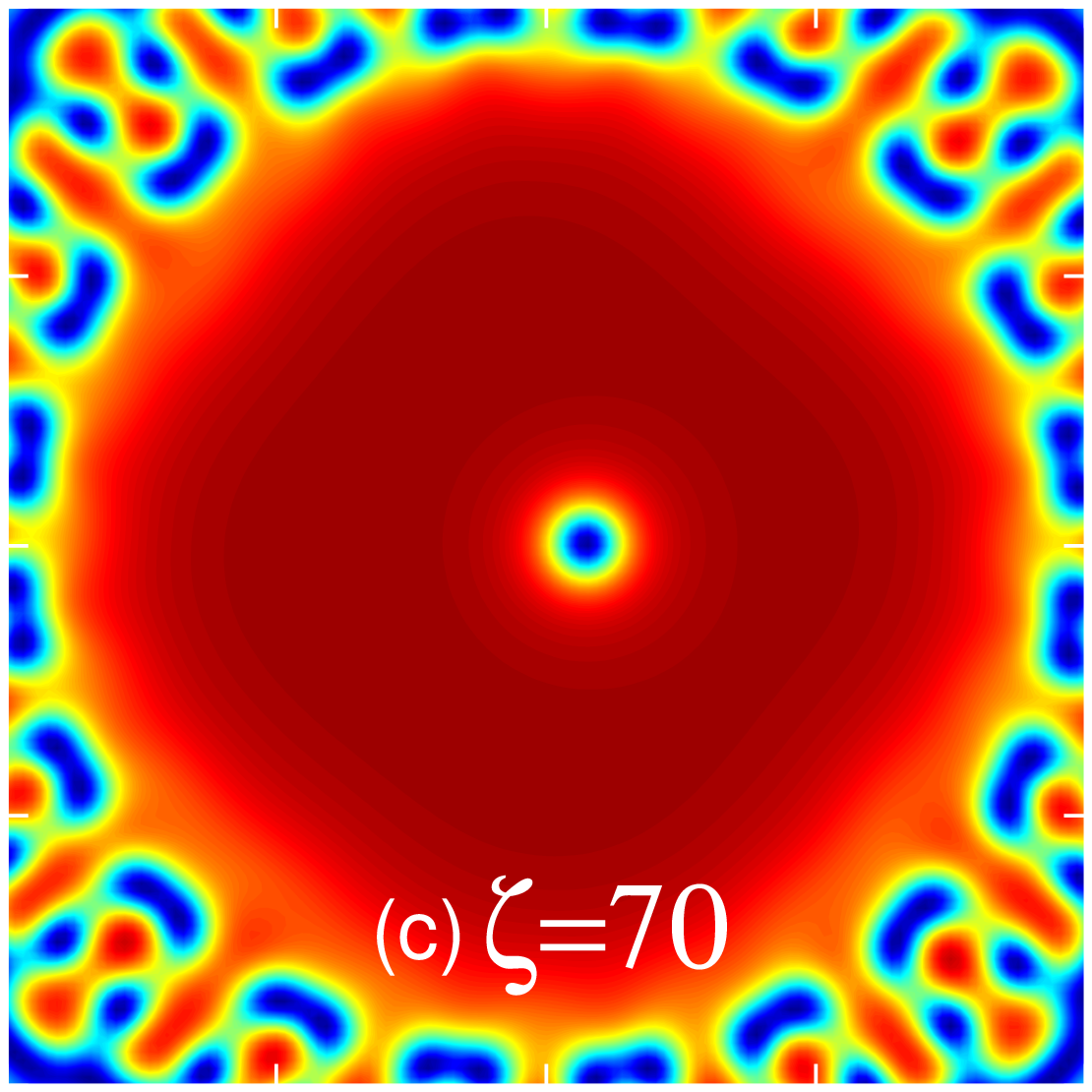}
\caption{\label{Fig5} A singly-charged OVS initially displaced $1.5$ from the center of the NBB background with $b_\infty=4.93$, having a central maximum of ${\rm FWHM}=38.4$, as the backgrounds in Fig. \ref{Fig1}, at the indicated propagation distances.}
\end{figure}

Our simulations show that OVSs embedded in flat NBBs survive for much longer distances than in Gaussian or SG backgrounds, and that the vortex dynamics is particularly simple, approaching that in the infinite background. Also, nesting the OVS in the NBB platform does not appreciably destabilize it, but instability initiates in the NBB periphery, allowing the quasi-ideal OVS dynamics to continue even if the background initiates to disintegrate. For example, the only difference in Fig. \ref{Fig5} with respect to Fig. \ref{Fig1} is that the background is a vortex-less NBB of the same FWHM and similar flatness as the SG background. The NBB-OVS system propagate without appreciable change up to $\zeta \sim 55$, while distortions were already present at $\zeta \sim 5$ with the SG background. At $\zeta=55$ the instability of the NBB starts to develop. A linear-instability analysis, as those reported in Refs. \cite{DES05,POR4}, reveals that vortex-less NBBs are unstable above a certain value of $C$, i. e., above a certain width of the central maximum. Also, the winding numbers of the unstable modes are increasingly high (e. g., $20$ for the most unstable mode of the NBB in Fig. \ref{Fig5}). Unstable modes of such a high winding number are highly delocalized and are not easily excited by perturbations inside the flat central maximum, e. g., by the nested OVS. Indeed instability is first observed in the periphery of the platform and first rings, endowing the periphery with a polygonal shape with a number of sides equal to the winding number of the dominant mode, but leaving its interior and the OVS substantially unaltered, as seen in Fig. \ref{Fig5}(b). In additional simulations we have confirmed that instability onsets at approximately the same distance with or without embedded vortex. As seen in Fig. \ref{Fig5}(c) for the regime of well-developed instability, the OVS still survives at rest. Of course, the ulterior development of instability leads to the destruction of the NBB-OVS sytem.

\begin{figure}[b]
\begin{center}
\includegraphics*[width=2.9cm]{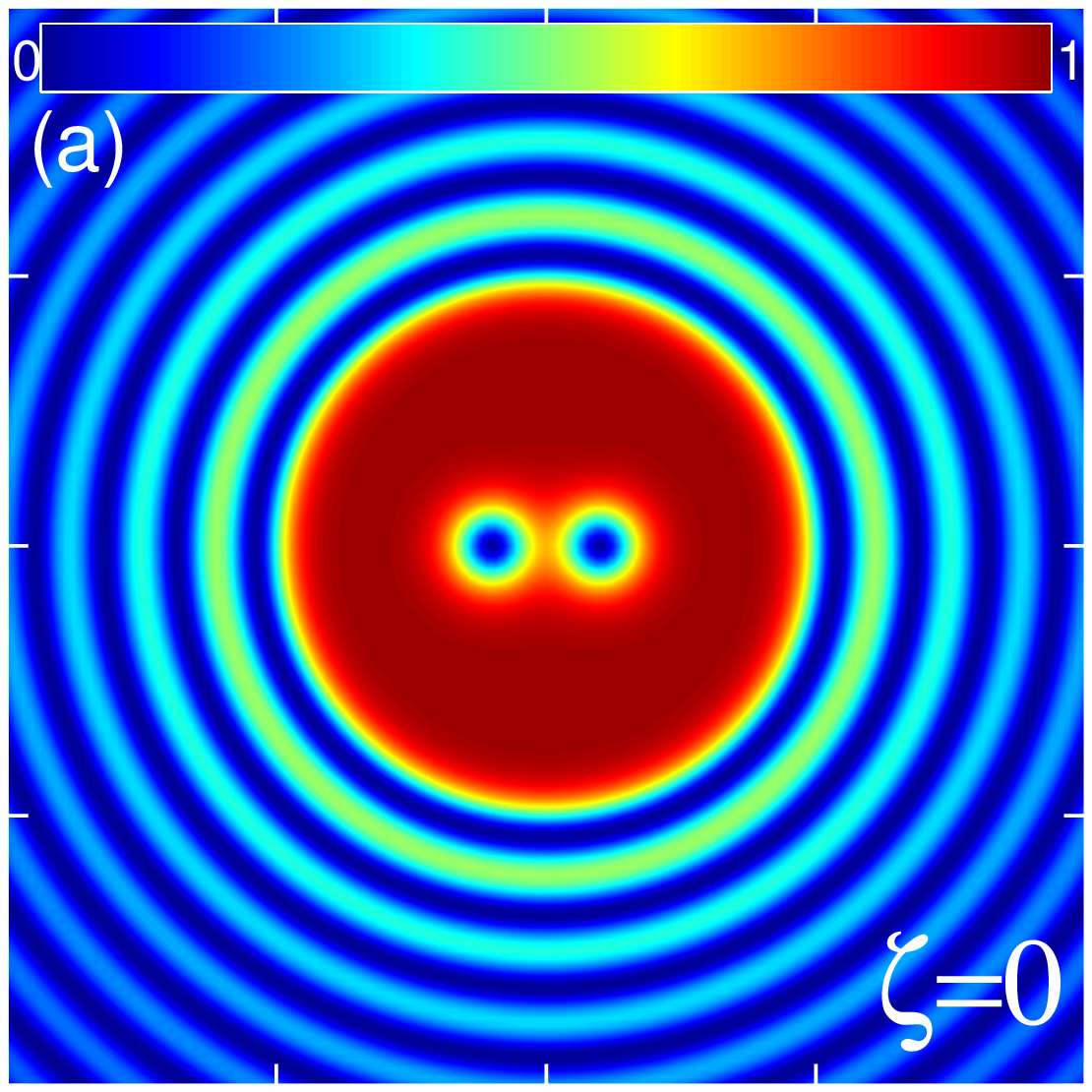}\includegraphics*[width=2.9cm]{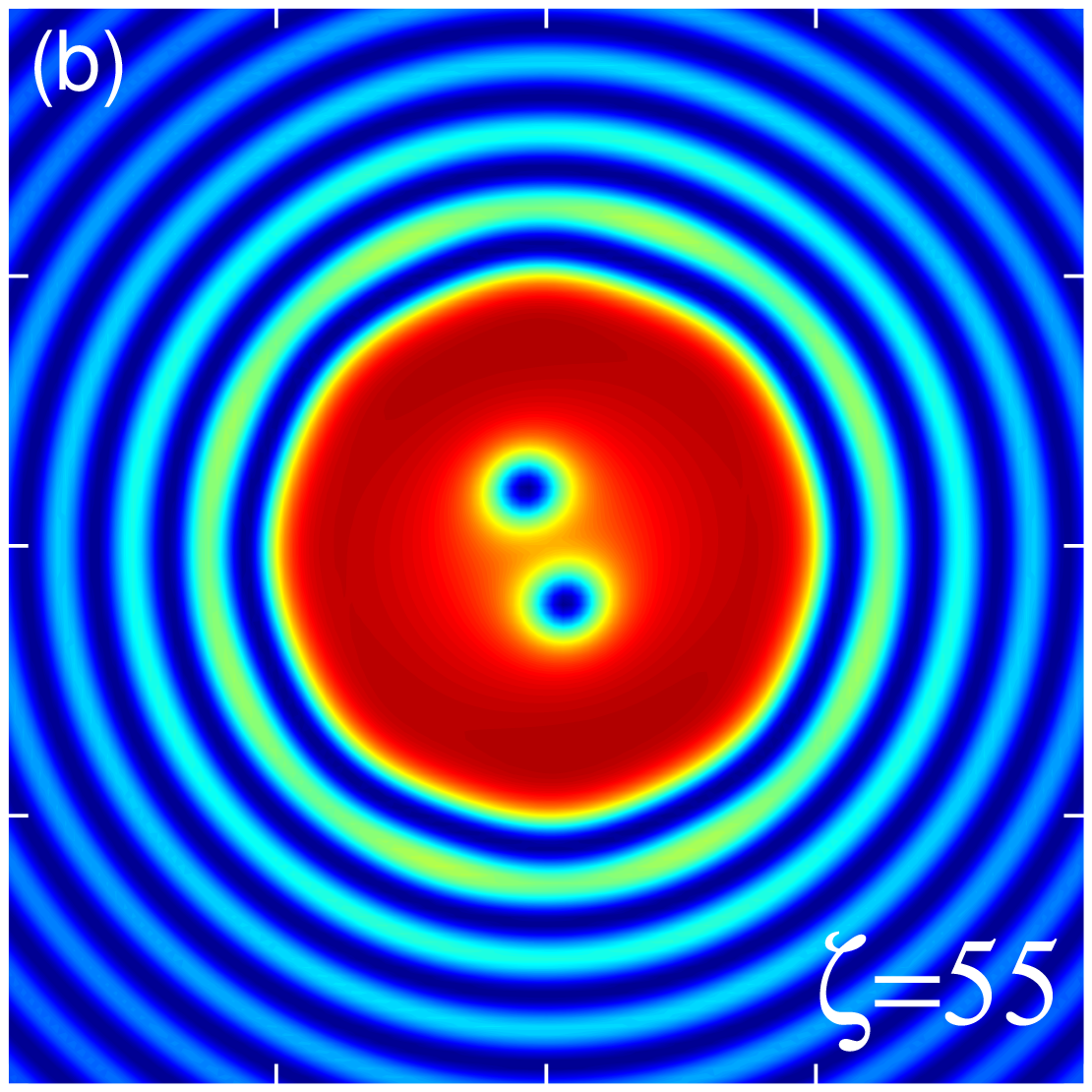}\includegraphics*[width=2.9cm]{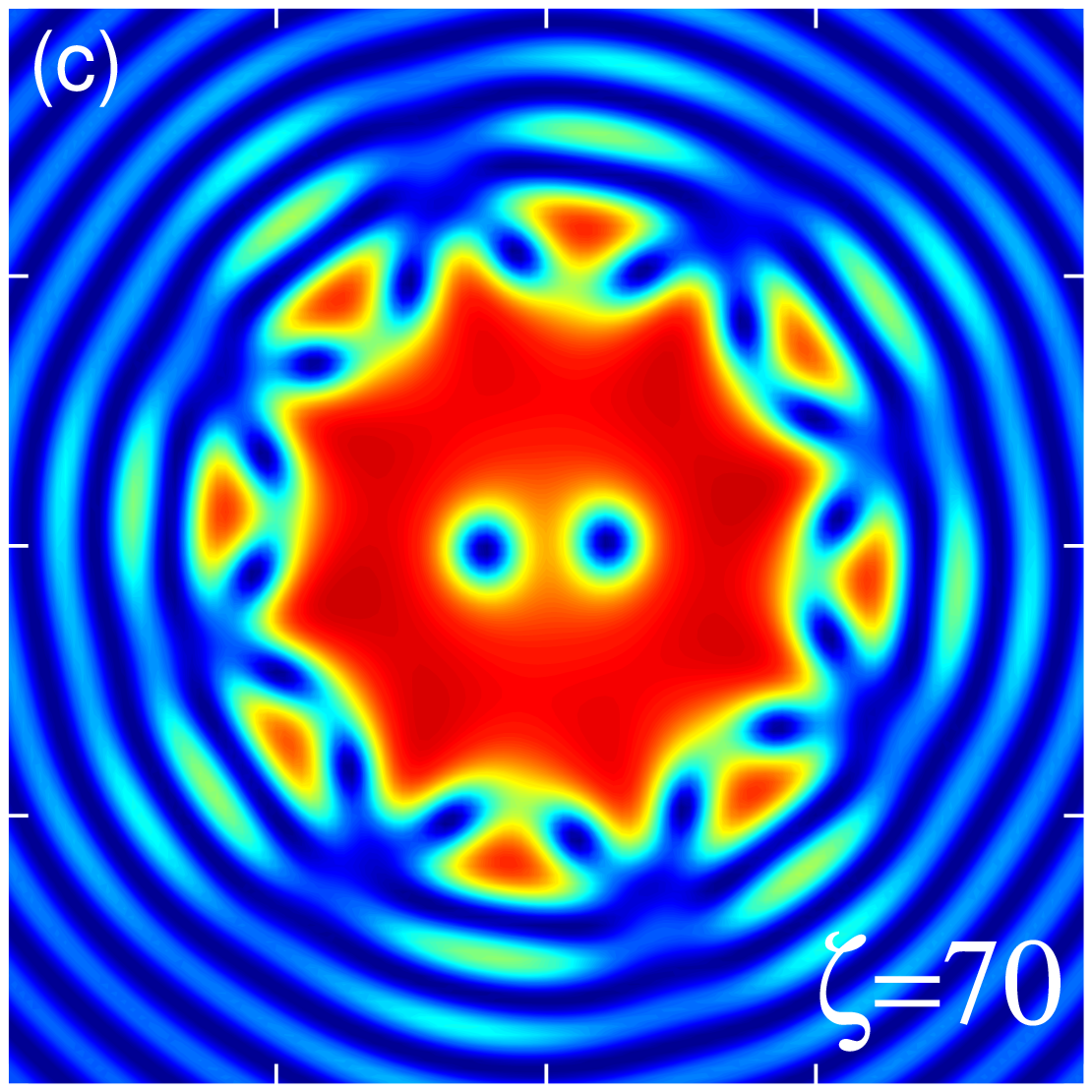}
\includegraphics*[width=4.3cm]{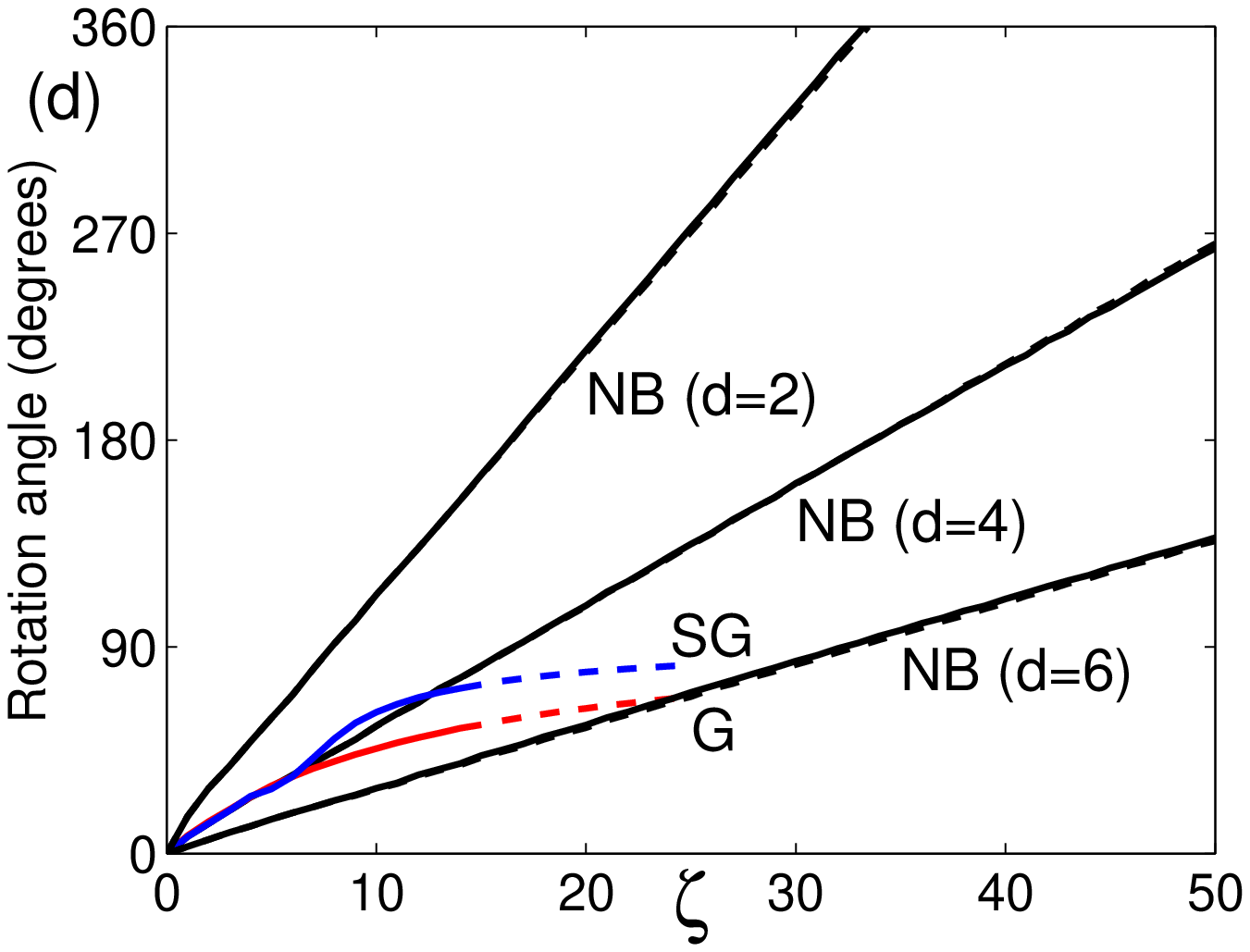}\includegraphics*[width=4.3cm]{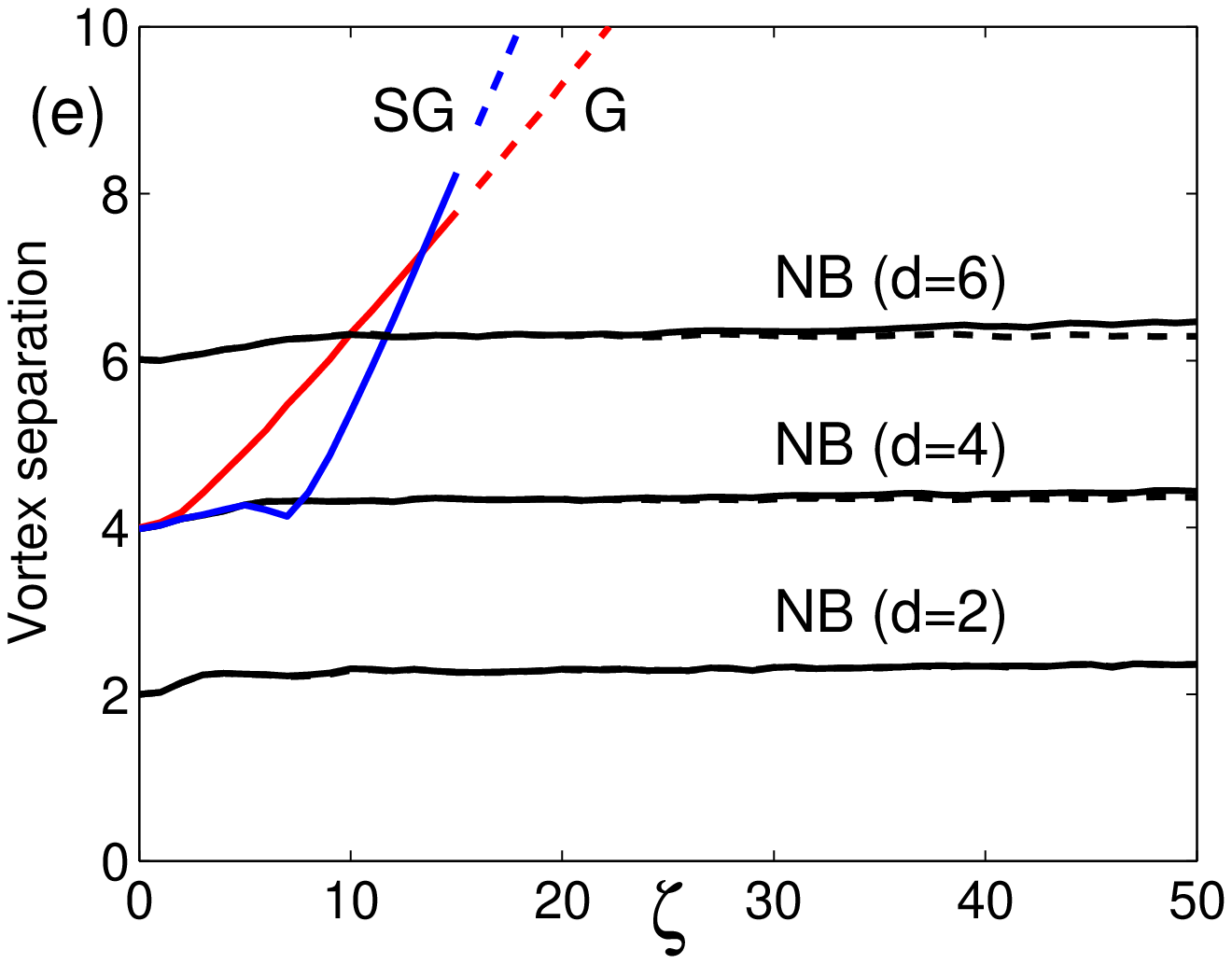}
\end{center}
\caption{\label{Fig6} (a-c) Two singly-charged OVSs initially separated $d=4$ in the fundamental NBB background with $C=0.99999999$, or $b_\infty=3.45$, and ${\rm FWHM}\simeq 20$, as the Gaussian and SG backgrounds in Fig. \ref{Fig2}, at the indicated propagation distances. (d) Rotation angle and (e) separation distance of two OVSs in the above NBB background as a function of propagation distance, for separations $d=2,4$ and $6$ (solid curves), compared to the same quantities in the uniform background and same separations (dashed curves). The red and blue curves are for the Gaussian and SG backgrounds and $d=4$.}
\end{figure}

Similarly, the only difference in Fig. \ref{Fig6}(a-c) with respect to Fig. \ref{Fig2}(c-f) is the NBB background, which however has the same FWHM. Instability of the NBB is weaker because of its narrower platform, but the two OVS are closer to its boundary. All together, the whole system propagates without appreciable distortion up to a similar distance $\zeta \simeq 55$ as in the preceding example, where instability arises as a weak octagonal distortion, and leads to complete destruction beyond $\zeta=70$. Importantly, the two OVSs are seen in Figs. \ref{Fig6}(a-c) to rotate at uniform angular velocity at a constant separation up to that distance, as in the ideal uniform background in Figs. \ref{Fig2}(a-b), while in the non-uniform backgrounds of Figs. \ref{Fig2} (c-f), broadening, spiraling and deceleration is already significant at $\zeta\sim 10$ \cite{NESH98}. This is further supported by Figs. \ref{Fig6}(d) and (e), where the dynamics of the two OVSs (rotation angle and OVS separation) placed initially at different distances ($d=2, 4$ and $6$) in the NBB and in the uniform background are seen to be almost identical, and therefore are not subjected to the inhomogeneities and spreading effects that complicates the vortex dynamics. In contrast to previous simulations and experiments, in which the maximum reported rotation is about $180^\circ$ (including significant broadening and spiraling), rotations larger than $360^\circ$ without deformation or distancing are easily attainable, as in Fig. \ref{Fig6}(c) for $d=4$ at $\zeta=70$ or for $d=2$ at $\zeta=33$ in Fig. \ref{Fig6}(d) (without background distortion this time).

In summary, we have reported the properties of non-diffracting, fundamental and high-order NBBs in self-defocusing media. The flat, arbitrarily wide intensity and phase profiles of the fundamental NBB makes it particularly suitable for nesting dark solitons. The vortices can survive unaltered for distances that are one order of magnitude larger than in wide Gaussian or SG, but quickly spreading backgrounds, and feature a particularly simple, and therefore predictable, dynamics and interactions, mimicking that in the ideal, unlimited  background. Further research is necessary to lengthen further the quasi-ideal propagation distance by suppressing the instability observed in the wide NBBs. This research has also revealed a close connection between high-order NBBs in self-defocusing media and OVSs: a OVS is the limiting NBB of infinitely wide first ring. From the opposite side, a high-order NBB can be regarded to be an OVS and its natural, finite (real) background. These results open new perspectives in applications such as optical vortex-induced waveguides, particle trapping and laser material processing, where precise steering of the OVSs is a crucial issue.

We acknowledge support from National Science Foundation (NSF) (1263236, 0968895, 1102301), The 863 Program (2013AA014402), and
Project of the Spanish Ministerio de Econom\'{\i}a y Competitividad No. MTM2015-63914-P.

\end{document}